\documentclass[prb, reprint, twocolumn, floatfix,amsmath,amssymb,apsrev,showpacs]{revtex4}%
\usepackage[colorlinks,urlcolor=blue,linkcolor=blue,anchorcolor=blue,citecolor=blue]{hyperref}
\usepackage{multirow}
\usepackage{color}
\usepackage{amsfonts}
\usepackage{amsmath}
\usepackage{mathrsfs}
\usepackage{amssymb}
\usepackage{wasysym}
\usepackage{xfrac}
\usepackage{graphicx}
\usepackage{graphics}
\usepackage{rotating}
\usepackage{lscape}

\begin{document}
%
\title{Quantum phase transition of the Bose-Hubbard model on cubic lattice \\with anisotropic hopping}
%
\author{Tao Wang}
\affiliation{Department of Physics, and Center of Quantum Materials and Devices,	Chongqing University, Chongqing, 401331, China}
%
\author{Xue-Feng Zhang}
\thanks{Corresponding author:zhangxf@cqu.edu.cn}
\affiliation{Department of Physics, and Center of Quantum Materials and Devices,	Chongqing University, Chongqing, 401331, China}

\begin{abstract}
In quantum many-body system, dimensionality plays a critical role on type of the quantum phase transition. In order to study the quantum system during dimensional crossover, we studied the Bose-Hubbard model on cubic lattice with anisotropic hopping by using the high order symbolic strong coupling expansion method. The analytic series expanded boundaries between the Mott-insulator and superfluid phase up to eighth order are calculated. The critical exponents are extracted by Pad\'{e} re-summation method, which clearly shows the dimensional crossover behavior. Meanwhile, the critical points at commensurate filling can also be obtained, and they match well with the prediction of renormalization group theory. The scaling of the gap energy and whole phase diagram are given at last, and they can be taken as the benchmark for experiment and numerical simulations in the future study.
\end{abstract}
\pacs{64.70.Tg,73.43.Nq,11.15.Me}
\maketitle

\section{Introduction}
Ultra-cold bosonic atoms trapped in the optical lattice \cite{Bloch_2002}, which is described by the Bose-Hubbard model \cite{Fisher_1989,Zoller_1998}, is considered as a good platform for quantum simulation due to its flexibility and fine-tunability of various quantum parameters \cite{Bloch_2008,Bloch_2017}. 
By interfering three laser beams which perpendicular to each others and their counter-propagating ones, the 3D optical lattice can be constructed, and the quantum phase transition (QPT) \cite{Sachdev_2007} between Mott-insulator and superfluid phase can be easily observed with time of the flight by tuning the strength of the laser beams (equivalent to depth of the lattice potential) \cite{Bloch_2002}. 
Meanwhile, the system in different dimensions can be realized by varying the strength of the laser beams in different directions \cite{Esslinger_04,Ott_2014}. As the statement of Mermin-Wagner theorem \cite{Mermin_1966}, the type of phase transition is highly related to the dimensionality. 
In one dimension, the Mott-insulator and superfluid QPT belongs to Berezinskii-Kosterlitz-Thouless (BKT) universality class, and it results in the re-entrance behavior near the tip of Mott lobe \cite{Esslinger_04,Monien_1998}. Moreover, the Bose-Hubbard model in higher dimensions ($d\ge2$) can be analytically solved with mean-field approach \cite{Fisher_1989}, and the QPT is numerically verified to be (d+1)D XY universality class \cite{Boris_2007,Boris_2008}.

Then, these motivate the research of dimensional crossover for ultra-cold bosons in optical lattice \cite{Esslinger_04,Giamarchi_2004,Giamarchi_2006,Bourbonnais_1991,Wang_2012,Schneider_2013,Pollet_2014,Ott_2014,Pelster_2015,Pelster_2017}, which can be easily experimentally realized by coupling arrays of $1$D bosonic tubes \cite{Schneider_2013}. Basing on the mean field theory, renormalization group (RG) approach \cite{Giamarchi_2004,Giamarchi_2006} and quantum Monte-Carlo (QMC) simulations \cite{Wang_2012,Pollet_2014}, for $1$D to $2$D dimensional crossover, the BKT type QPT only exists in exactly one dimension. It means infinitesimal inter-site tunnelling in other directions will turn the transition to be 3D XY type in the thermodynamic limit \cite{Larkin_1974}. In addition, the changing of critical points at the commensurate filling are numerically demonstrated to be well described by RG method\cite{Giamarchi_2004,Giamarchi_2006,Bourbonnais_1991,Boris_2016}. However, it is difficult to numerically study the QPT of $1$D to $3$D dimensional crossover, because of drawbacks of different numerical approaches\cite{Numerical_2013}.

In this paper, we use high order symbolic strong coupling expansion method (HSSCE) \cite{Monien_1996,Eckardt_2009,Zhang_2018} to study $2$D arrays of coupled $1$D tubes of ultra-cold bosonic gas, where the inter-tube coupling is varied from zero to the same value as the intra-tube coupling, so that the system can be detuned from $1$D to $3$D. Basing on the process-chain algorithm \cite{Eckardt_2009}, strong coupling expansion \cite{Monien_1996} and symbolization, the HSSCE can give the symbolic series expansion function of the critical line between compressible phase (Mott-insulator) and incompressible one (superfluid) up to very high order for different atom filling number and anisotropy (up to eighth order in this work list in Appendix.\ref{AA}). In previous work \cite{Zhang_2018}, we implement HSSCE to isotropic $1$D, $2$D and $3$D Bose-Hubbard model, together with Pad\'{e} re-summation, we get very accurate phase boundaries, which are non-distinguishable from numerical results, and also the critical exponents obtained from the function of charge gap. Thus, we expect the HSSCE can also provide the high accuracy results of the anisotropic 3D Bose-Hubbard model, which is not only benefit for understanding the QPT during the 1D to 3D dimensional crossover, but also used for benchmark of the experiments and numerical simulations.


This paper is structured as follows. In Sec.\ref{model}, after briefly introducing the anisotropic 3D Bose-Hubbard model, we describe how to implement the HSSCE on this model, especially dealing with the anisotropy. Then, by listing the eighth order symbolic results in Appendix \ref{AA}, we discuss the universality class of phase transition changing during the 1D-3D dimensional crossover in Sec.\ref{result}, and give the quantum phase diagram of Mott-insulator to superfluid phase for different filling and anisotropy. Meanwhile, we demonstrate the tip of Mott lobe fulfills the RG prediction for different filling. At last, we make a conclusion in  Sec.\ref{conclusion}.
\section{Model and Method}\label{model}
The model we considered is the Bose-Hubbard model on cubic lattice with anisotropic hopping amplitude:
\begin{eqnarray}
\hat{H}&=&- t\left(\sum_{<i,j>_x}\hat{b}_{i}^{\dagger}\hat{b}_{j}+\alpha \sum_{<i,j>_{y,z}}\hat{b}_{i}^{\dagger}\hat{b}_{j}+h.c.\right) \nonumber\\
&&+\frac{U}{2}\sum_i \hat{n}_i(\hat{n}_i-1) 
-\mu\sum_i\hat{n}_i,
\end{eqnarray}
where the hopping amplitudes in $x$ direction is $t$ and in $y,z$ directions are $\alpha t$ with the anisotropic parameter $0\le \alpha\le 1$; $\hat{b}_i^{\dagger}$ ($\hat{b}_i$) is the creation (annihilation) operator on site $i$; $\hat{n}_i$ is the local density operator; $U$ is the on-site two-body interaction which can be detuned with Feshbach resonance and $\mu$ is the chemical potential. When $\alpha$ is equal to one, the system is isotropic in three dimension. While in another limit $\alpha=0$, all the hoppings in $y$ and $z$ directions are shut off and 2D array of Bose-Hubbard chains are decoupled, so that the system can be treated as $1$D. In both limits, there exist the QPTs between Mott-insulator and superfluid phase. The upper (lower) phase boundary of the Mott lobe is caused by particle (hole) excitation, and the dynamical exponent is $z=2$, while at the tip of lobe, the effective particle-hole symmetry leads to $z=1$ \cite{Fisher_1989}. The whole phase diagrams can be obtained with degenerate perturbation theory named strong coupling expansion method \cite{Monien_1996}.

Taking the hopping matrix as the perturbative term, the strong coupling expansion method aims to calculate the energies of the ground state and single-particle (hole) excited state in the Mott-insulator. Then, the series expansion function of the upper (lower) phase boundary can be obtained by comparing the ground state energy with single-particle (hole) exited one. However, the exponential increment of the calculation for higher orders prevents reaching high precision.

The process chain approach \cite{Eckardt_2009} takes advantages of the Kato representations to diagram the perturbative contributions in each order. After simplifying the diagrams by considering the different symmetries (e.g. point symmetry) and topology, the number of diagrams can be significantly decreased. Then, we can achieve very high order (e.g. 10th order \cite{Zhang_2018}) by numerically computing each diagram. Meanwhile, the whole process is symbolized in our previous work \cite{Zhang_2018} so that the results can be considered as exact Taylor expansion of the phase boundaries, so it can provide more accurate analysis of the QPT. Considering the computational resource consumption is more sensitive to the orders than dimensionality, the HSSCE is very suitable for quantitatively studying the QPT during the dimensional crossover. 

The algorithm and source code of HSSCE are described and given in previous paper \cite{Zhang_2018}. Extending HSSCE to anisotropic system is pretty straight forward, the difference is simplification of the diagrams. Due to the anisotropy of the system, the point symmetry is lower than before, so the numbers of the equivalent diagrams are less than before. Thus, the calculation requires more computing hours than isotropic case. After comparing the perturbative energy of ground state and single-particle (hole) excited state, we obtain the upper and lower boundaries:
\begin{eqnarray}
\nonumber
{\rm particle:}&~&~\frac{\mu_{p}}{U}=n -\mathop{\sum}_k
\beta_{u}^{(k)}(\alpha,n) \left(\frac{t}{U}\right)^k,\\
{\rm hole:}&~&~\frac{\mu_{h}}{U}=n-1+\mathop{\sum}_k
\beta_{d}^{(k)}(\alpha,n) \left(\frac{t}{U}\right)^k, \label{eqboundary}
\end{eqnarray}
where the coefficient of boundaries are $\beta_{u(d)}^{(k)}(\alpha,n)=\sum_{i=1}^{k+1}\sum_{j=1}^{k+1}\alpha^{i-1} B^{u(d)}_{i,j}(k)n^{j-1}$ with the coefficient matrix $B^{u(d)}_{i,j}(k)$ which can be symbolically output by HSSCE and listed in Appendix.\ref{AA} up to eighth order.

\section{Results}\label{result}
The HSSCE is a symbolic algorithm, as presented in Eqn.\ref{eqboundary} and the Appendix.\ref{AA}, the results can be considered as the analytic Taylor expansion of boundaries for all the parameters. However, the boundaries have the singularities exist at the tip of Mott lobe. It means that only  the phase transition in the strongly coupled region $t/U\ll 1$ can be well described by the raw data, but not around the singularities. However, as the most important part of the QPT, the universality class can be well detected by checking the scaling behaviour of the gap energy $\Delta=\mu_p-\mu_h$ with respect to the quantum parameter $t/U$ around the tip of lobe. Thanks to the Pad\'{e} re-summation method, the behaviour of the singularity can be reappeared. In one dimension, the re-entrance behaviour demonstrates the BKT universality class, and the critical exponents $z\nu$ can be obtained for higher dimension in which the phase transition belong to the (d+1)D XY universality class \cite{Zhang_2018}. 

For the anisotropic system at finite temperature, there exists the dimensional phase transition from the 1D Luttinger liquid to the 3D Bose liquid \cite{Ott_2014}. In comparison, at zero temperature, the condensation can exist for arbitrary small anisotropic parameter $\alpha$, so the QPT for 1D-3D dimensional crossover region should be belong to $4$D XY type except pure 1D point $\alpha=0$ which belong to BKT type \cite{Larkin_1974}. 
\begin{figure}[t]
	\includegraphics[width=0.5\textwidth]{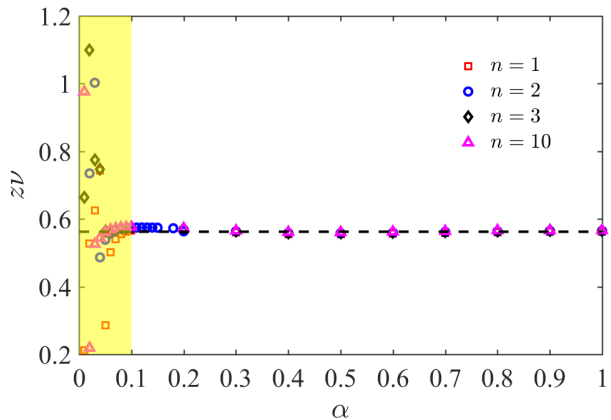}
	\caption{The critical exponents $z\nu$ changing with $\alpha$ for different filling number $n$.  For most of the region, the values of $z\nu$ are approximately around $0.56$ (black dash line) and less sensitive to the anisotropy $\alpha$ and filling number $n$. However, When $\alpha\apprle0.1$ (yellow region), the critical exponents $z\nu$ become messy.}\label{fig01}
\end{figure}

In the thermodynamic limit, the charge gap of Bose-Hubbard model in d$\ge2$ dimension follows the power law scaling $\Delta=A(t)(t_c-t)^{z\nu}$ \cite{Fisher_1989}, where $A(t)$ is a regular function and $z,\nu$ denotes the critical exponents which characterize the scaling of long range correlation. Thus, the logarithm of charge gap 
\begin{equation}
\frac{\partial log\Delta}{\partial t}=\frac{A^{'}(t)}{A(t)}+\frac{z\nu}{t_c-t}
\end{equation}
has a simple pole $t_c$ and could be determined by Pad\'{e} re-summation method:
\begin{equation}
\frac{\partial log\Delta}{\partial t}=\frac{\sum_{n=0}^{n_{max}}a_nt^n}{1+\sum_{m=0}^{m_{max}}b_mt^m}.\label{pade}
\end{equation}
The coefficients $a_n$ and $b_m$ could be calculated by fitting with the restriction $n_{max}+m_{max}=8$. Here we choose the most nature way $n_{max}=m_{max}=4$ (other possible choices are discussed in Appendix. \ref{AA2}). The smallest real positive simple pole gives the critical hopping amplitude $t_c$ at the lobe tip or commensurate fillings, while the corresponding residue yields the critical exponent $z\nu$.

We show the critical exponents $z\nu$ for different filling and anisotropy in Fig.\ref{fig01}. When $\alpha \apprge 0.1$, the results demonstrate the phase transitions are of 4D XY type universality class with critical exponents same as the isotropic 3D case $z\nu \approx 0.5625$. Because $d+1=4$ is the upper critical dimension of the XY model, the logarithm correction may result in the discrepancy with the meanfield value $z\nu=1/2$. However, when approaching to the 1D limit, the fitting of critical exponents start to be invalid. Such phenomena are not related to the filling numbers, and we think they result from the limitation of perturbative orders. As shown in Appendix.\ref{AA}, for each perturbative term, each element of coefficient matrix are also series expanded with respect to $\alpha$. Thus, when $\alpha$ is less than $0.1$, the high power order of $\alpha$ will contribute less, and it means the result is more like the 1D case. To make the 4D XY type fitting still work for strong anisotropy, the higher order terms are necessary. 

\begin{figure}[t]
	\includegraphics[width=0.5\textwidth]{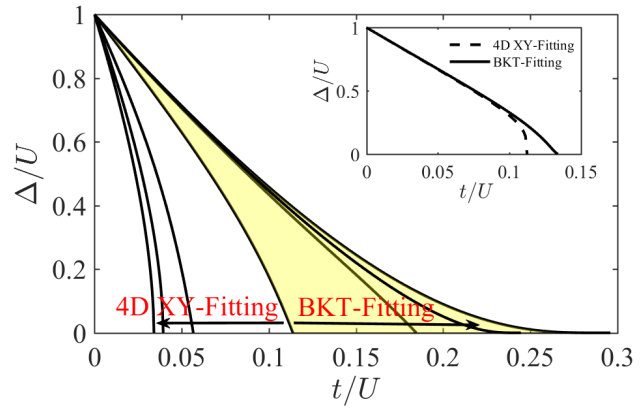}
	\caption{The charge gap function with different $\alpha$ (right to left: 0, 0.005, 0.01, 0.1, 0.5, 0.8 and 1) for $n=1$. For $\alpha\ge0.1$, the 4D XY type fitting $A(t)(t_c-t)^{z\nu}$ is adopted, while $\alpha<0.1$, the BKT type fitting  $B(t)\exp\left(-\frac{W}{\sqrt{t_c-t}}\right)$ is chosen. Inset: Two types of the fitting at $\alpha=0.05$ and $n=1$.}\label{fig02}
\end{figure}

In one dimension, the Bose-Hubbard chain undergoes the BKT QPT, and the charge gap fulfils exponential scaling behaviour  $\Delta=B(t)\exp\left(-\frac{W}{\sqrt{t_c-t}}\right)$, where $B(t)$ is a regular function, $W$ is the fitting parameter and $t_c$ is the critical point which could also be captured by Pad\'{e} re-summation of $[\log\Delta(t)]^2$. Different from the 4D-XY type fitting, the smallest real positive pole is taken after cancelling out the common zero points of denominator and numerator. As shown in the inset of Fig.\ref{fig02}, when $\alpha=0.05$ which is smaller than the critical value $0.1$, the charge gap function with 4D XY-type fitting exhibits a rounded tip which highly conflicts with the sharp shape in one dimensional system. In order to depict the whole anisotropy region in a proper way, as shown in Fig.\ref{fig02}, we adopt 4D XY-type and BKT-type fitting for $\alpha\ge0.1$ and $\alpha<0.1$, respectively. We can find the charge gap function changing from rounded to sharp when decreasing $\alpha$, and it demonstrates the scaling behaviour changing during the dimensional crossover. If higher order terms can be considered in the future work, we think the deep anisotropic region can be approached with 4D XY-type fitting.
Considering the high accuracy of the HSSCE, our results of charge gap function can be taken as benchmark for numerical simulations \cite{Monien_1998,Boris_2007,Boris_2008} and experiments \cite{Esslinger_04,Bloch_2002}.

The anisotropic 3D system can be treated as 2D array of coupled 1D tube. Because the one dimensional Bose-Hubbard model can be well described by the bosonization \cite{Sebastian_2009}, the dimensional crossover problem can be analytically solved with help of RG approach \cite{Giamarchi_2004,Giamarchi_2006,Boris_2016,Sebastian_2017}. After taking couplings in $y (z)$ directions as the perturbation of the effective bosonized 1D Hamiltonian, the RG equations can be constructed \cite{Giamarchi_2004,Giamarchi_2006,Boris_2016,Pollet_2014,Esslinger_04,Ott_2014} as follows:
\begin{eqnarray}
\frac{dg_J}{dl}&=&(2-\frac{1}{2K})g_J\\
\frac{dg_u}{dl}&=&(2-K)g_u\\
\frac{dK}{dl}&=&4g_J^2-g_u^2K^2, \label{RG}
\end{eqnarray}
where $g_J$ and $g_u$ are dimensionless parameters characterize the strength of hopping in y(z) direction and periodical `Mott potential', $K$ is the Luttinger parameter of 1D Bose-Hubbard chain and $l$ is the flow parameter. After comparing with the RG equations of 1D-2D crossover \cite{Pollet_2014}, we can find the only difference is that the coefficient of first term in $dK/dl$ is $4$ not $2$. Because the RG analysis is around $K(l)\approx 2$ and $g_J$ is set to be zero,  the analytical relation between anisotropy $\alpha$ and critical point $t_c(\alpha)$ for the superfluid to Mott-insulator transition at commensurate fillings is same:
\begin{equation}
\alpha=C\exp(-\frac{\pi s}{4b\sqrt{t_c^{1d}/t_c(\alpha)-1}}),\label{scale}
\end{equation}
in which $C$ and $b$ are constants, $t_c^{1d}$ is the critical hopping amplitude of the lobe tip for $1$D Bose-Hubbard model, while $s$ is the constant related to the scaling of order parameter. Because the charge gap $\Delta\propto 1/\xi$ where $\xi$ stands for correlation length, here the constant $s$ is set to be $1$. Noting that, different from Ref.\cite{Pollet_2014}, we set $U=1$ as energy unit.

\begin{figure}[t]
	\includegraphics[width=0.5\textwidth]{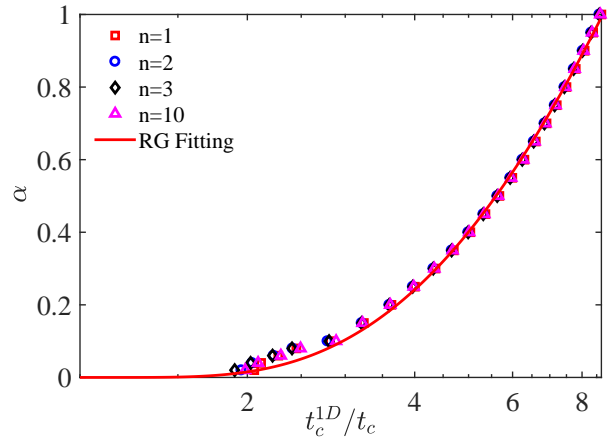}
	\caption{The relation between critical hopping amplitude $t_c$ at commensurate fillings and the anisotropic parameter $\alpha$. The discrete data point are gotten from Pad\'{e} re-summation Eqn.\ref{pade} for different filling number. The red line is the theoretical fitting basing on Eqn.\ref{scale}. }\label{fig03}
\end{figure}

As shown in Fig.\ref{fig03}, similar to the 1D to 2D dimensional crossover \cite{Pollet_2014}, the HSSCE data of critical points at the tip of Mott-lobes for different filling numbers nearly coincide with the fitting curve with Eqn.\ref{scale} except the small deviations when $\alpha$ is small. The reason for these deviations is because we use BKT-type fitting for small $\alpha$ which is not consistent with the 4D XY-type universality class concluded by RG theory. Such mismatch can be recovered by including higher order in the future work. The value of constants $C$ and $b$ are list in Table \ref{constants}, and they are less relevant to the filling number. Furthermore, we find the fitting parameter $b$ is very closed to the 1D to 2D dimensional crossover from QMC simulation \cite{Pollet_2014}.
\begin{table}[h]
	\begin{center}
		\begin{tabular}{|c|c|c|c|c|}
			\hline
			constants & $n=1$ & $n=2 $ &$n=3$ & $n=10$ \\
			\hline
			$C$& $10.536$ & $10.460$ &  $10.455$ &$10.482$   \\
			\hline
			$b$ & $0.120$ & $0.121$ & $0.121$ &$0.120$  \\
			\hline
		\end{tabular}
		\caption{The fitting constants for different filling number basing on the Eqn. \ref{scale}.}\label{constants}
	\end{center}
\end{table}

\begin{figure}[h]
	\includegraphics[width=0.5\textwidth]{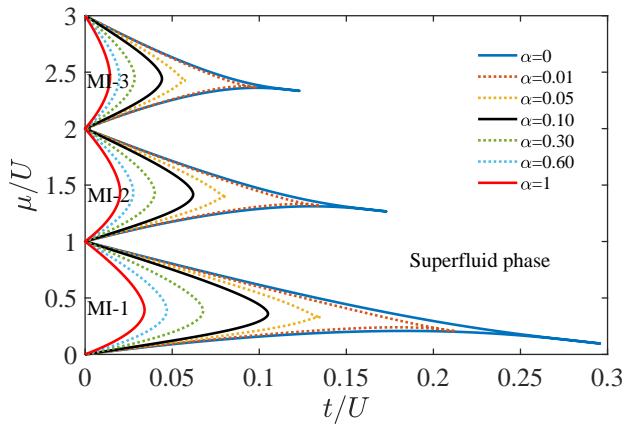}
	\caption{The whole phase diagram of the anisotropic Bose-Hubbard model for different anisotropy $\alpha$ (right to left: 0, 0.01, 0.05, 0.10, 0.30, 0.60, 1.00) and different fillings.}\label{fig04}
\end{figure}

At last, we discuss the whole phase diagram at incommensurate fillings presented in Fig.\ref{fig04}. Because the HSSCE method gives the series expansion results of upper and lower boundaries, the phase diagram in the grand canonical ensemble are explicit provided. As discussed in our previous paper \cite{Zhang_2018}, in three dimension, all coefficients in different orders are positive, so that there is no exotic behavior observed in the phase diagram. In comparison, in one dimensional system, some of the coefficients are changed to negative, so that three critical points could exist at same chemical potential energy $\mu/U$ which is called re-entrance behavior. Such exotic phenomena are due to the exponential decay of the gap energy and also reflects the QPT is belong to BKT universality class. In Fig.\ref{fig04}, the phase boundaries are changing from rounded shape to the sharp one. Meanwhile, the re-entrance also can be found for small $\alpha$ which reflects that the dimensional crossover becomes serious when approaching to one dimension.

\section{Conclusion and discussion}\label{conclusion}
Taking advantage of the HSSCE method, we get the symbolic series expanded phase boundaries between Mott-insulator and superfluid phase for anisotropic Bose-Hubbard model on cubic lattice up to eighth order. With Pad\'{e} re-summation method, the related critical exponents $z\nu$ can be extracted. However, when anisotropy is strong $\alpha<0.1$,  $z\nu$  changes drastically. In order to restore the physical process, we reconstruct the phase boundaries by choosing 4D XY type fitting for $\alpha\ge 0.1$, and BKT type for $\alpha<0.1$. Then, we studied the relation between the critical point at commensurate filling and the anisotropy, and find it matches well with the prediction given by RG theory. At last, we also give the whole phase diagram at incommensurate filling for different anisotropy $\alpha$.

Because of the high flexibility and tunability, the results presented in this work could be checked by cold atom gas experiments in optical lattice. One could prepare $1$D Boson gases trapped in $2$D optical lattice\cite{Ott_2014}, thus the hopping amplitude between tubes could be tuned by intensity of laser coupling the tubes. The phase transition between superfluid and Mott insulator could be captured by time-of-flight technical. The location of the tip lobe as the function of anisotropic parameter could be gotten by detecting the point where plateaus begin from the expansion experiment\cite{Schneider_2013}. In addition, when the trapping potential is introduced, the system will exhibit the pancake structures. The phase boundaries can be extracted from the region of each Mott plateau, and can be compared with our results \cite{Svistunov_2010}. 

\begin{acknowledgements}
We are thankful for useful discussions with  Axel Pelster, Sebastian Eggert, Hong-Hao Tu and Long Liang.  X.-F. Z. acknowledges funding from the National Science Foundation of China under Grants No. 11804034, No. 11874094 and No. 11947406, and Grant No. cstc2018jcyjAX0399 by Chongqing Natural Science Foundation.  T. W. acknowledges funding from China Postdoctoral Science Foundation funded project 2020M673118.
\end{acknowledgements}

\clearpage
\begin{appendix}
\begin{widetext}
\section{The symbolic series expansion of the phase boundaries for anisotropic Bose-Hubbard model} \label{AA}
The boundaries of Mott lobe can be represented with series expansion form:
\begin{eqnarray}
\nonumber
{\rm particle:}&~&~\frac{\mu_{p}}{U}=n -\mathop{\sum}_k
\beta_{u}^{(k)}(\alpha,n) \left(\frac{t}{U}\right)^k,\\
{\rm hole:}&~&~\frac{\mu_{h}}{U}=n-1+\mathop{\sum}_k
\beta_{d}^{(k)}(\alpha,n) \left(\frac{t}{U}\right)^k, 
\label{A1}
\end{eqnarray}
where the coefficient of upper boundary is $\beta_{u}^{(k)}(\alpha,n)=\sum_{i=1}^{k+1}\sum_{j=1}^{k+1}\alpha^{i-1} B^{u}_{i,j}(k)n^{j-1}$ with the coefficient matrix $B^{u}_{i,j}(k)$ are
\begin{eqnarray*}
	\textbf{B}^{u}(1)&=&\left(
	\begin{array}{cc}
		-4 & -4 \\
		-2 & -2 \\
	\end{array}
	\right),\hspace{10px}
	\textbf{B}^{u}(2)=\left(
	\begin{array}{ccc}
		0 & -8 & -6 \\
		0 & -16 & -16 \\
		0 & 0 & 1 \\
	\end{array}
	\right),\hspace{10px}
	\textbf{B}^{u}(3)=\left(
	\begin{array}{cccc}
		0 & -16 & -60 & -44 \\
		0 & -68 & -210 & -142 \\
		0 & -20 & -66 & -46 \\
		0 & 2 & 3 & 1 \\
	\end{array}
	\right),\\
	\textbf{B}^{u}(4)&=&\left(
	\begin{array}{ccccc}
		0 & -\frac{135}{2} & -\frac{17857}{60} & -\frac{2449}{6} & -\frac{10597}{60} \\
		0 & -316 & -1838 & -3036 & -1514 \\
		0 & -\frac{622}{3} & -\frac{15659}{15} & -\frac{4766}{3} & -\frac{11239}{15} \\
		0 & -\frac{92}{3} & -\frac{542}{3} & -\frac{892}{3} & -\frac{442}{3} \\
		0 & -\frac{9}{4} & -\frac{559}{120} & -\frac{7}{12} & \frac{221}{120} \\
	\end{array}
	\right),
	\textbf{B}^{u}(5)=\left(
	\begin{array}{cccccc}
		0 & \frac{31}{3} & -\frac{387253}{450} & -\frac{1638443}{450} & -\frac{2188403}{450} & -\frac{941863}{450} \\
		0 & -\frac{4051}{4} & -\frac{19097321}{1800} & -\frac{58121101}{1800} & -\frac{68366371}{1800} & -\frac{27519641}{1800} \\
		0 & -903 & -\frac{4684909}{450} & -\frac{14775229}{450} & -\frac{17720459}{450} & -\frac{7223789}{450} \\
		0 & -\frac{2522}{9} & -\frac{260017}{75} & -\frac{2510981}{225} & -\frac{1013317}{75} & -\frac{1245971}{225} \\
		0 & -\frac{209}{36} & -\frac{401707}{1800} & -\frac{1456567}{1800} & -\frac{1803257}{1800} & -\frac{737947}{1800} \\
		0 & \frac{653}{72} & \frac{114703}{3600} & \frac{42481}{1200} & \frac{56453}{3600} & \frac{11063}{3600} \\
	\end{array}
	\right),\\
	\textbf{B}^{u}(6)&=&\left(
	\begin{array}{ccccccc}
		0 & -\frac{6344897}{6048} & -\frac{12245886137}{1512000} & -\frac{695777273}{25200} & -\frac{70292096771}{1512000} & -\frac{5679281509}{151200} & -\frac{4396126631}{378000} \\
		0 & -\frac{510985}{108} & -\frac{3957402269}{63000} & -\frac{26732808157}{94500} & -\frac{26397560641}{47250} & -\frac{3942882856}{7875} & -\frac{31526579237}{189000} \\
		0 & -\frac{1597247}{216} & -\frac{13883931061}{162000} & -\frac{38573963749}{108000} & -\frac{434792528659}{648000} & -\frac{62651400559}{108000} & -\frac{24361535093}{129600} \\
		0 & -\frac{1029677}{252} & -\frac{1136383799}{21000} & -\frac{92334825139}{378000} & -\frac{40602475367}{84000} & -\frac{23428854167}{54000} & -\frac{36476848183}{252000} \\
		0 & -\frac{897971}{756} & -\frac{6445354697}{567000} & -\frac{10942072933}{252000} & -\frac{50536206163}{648000} & -\frac{49639768249}{756000} & -\frac{3794314171}{181440} \\
		0 & -\frac{4663}{54} & -\frac{53282041}{94500} & -\frac{98446741}{47250} & -\frac{186941341}{47250} & -\frac{9228077}{2625} & -\frac{12233309}{10500} \\
		0 & -\frac{329201}{12096} & -\frac{6145973}{51840} & -\frac{1299931}{7000} & -\frac{1019369927}{9072000} & -\frac{26507951}{1512000} & \frac{2100869}{4536000} \\
	\end{array}
	\right),\\
	\textbf{B}^{u}(7)&=&\left(
	\begin{array}{cccccc}
		0 & \frac{17208115}{5292} & \frac{29226199817}{8505000} & -\frac{40776313677359}{476280000} & -\frac{2860644143681}{7620480} & -\frac{17569043110187}{27216000} \\
		0 & -\frac{2820543409}{254016} & -\frac{261543252131069}{952560000} & -\frac{18123177499069}{9720000} & -\frac{10677162089359543}{1905120000} & -\frac{16028030330232943}{1905120000} \\
		0 & -\frac{541247299}{47628} & -\frac{7661308452911}{17860500} & -\frac{182365471982893}{57153600} & -\frac{28505938773306611}{2857680000} & -\frac{43743064347892991}{2857680000} \\
		0 & -\frac{832739197}{54432} & -\frac{586905879598571}{1428840000} & -\frac{8185350354028231}{2857680000} & -\frac{7126876800312161}{816480000} & -\frac{75531105849692267}{5715360000} \\
		0 & -\frac{642757627}{190512} & -\frac{93304063975439}{714420000} & -\frac{21387397577611}{22325625} & -\frac{4232810624737243}{1428840000} & -\frac{6444199749517843}{1428840000} \\
		0 & \frac{488721839}{381024} & -\frac{222894050851}{40824000} & -\frac{10436593904551}{114307200} & -\frac{1909402585897307}{5715360000} & -\frac{63304570520783}{116640000} \\
		0 & \frac{202502513}{762048} & \frac{2243122165069}{2857680000} & -\frac{194179244813}{178605000} & -\frac{47163130806559}{5715360000} & -\frac{83609709293119}{5715360000} \\
		0 & \frac{137725031}{1524096} & \frac{2608966642123}{5715360000} & \frac{2599569516137}{2857680000} & \frac{10042635987533}{11430720000} & \frac{4617015079493}{11430720000} \\
	\end{array}
	\ldots\ldots\right.\\
	&&\left.\ldots\ldots
	\begin{array}{cc}
		-\frac{482685930471859}{952560000} & -\frac{20517348363031}{136080000} \\
		-\frac{2354998048684739}{381024000} & -\frac{3373911641427409}{1905120000} \\
		-\frac{4656897396054443}{408240000} & -\frac{9436217113175831}{2857680000} \\
		-\frac{55815885881231213}{5715360000} & -\frac{3212480565675379}{1143072000} \\
		-\frac{136403063466893}{40824000} & -\frac{1375712097341929}{1428840000} \\
		-\frac{2369567032887617}{5715360000} & -\frac{99107423249021}{816480000} \\
		-\frac{64070255378107}{5715360000} & -\frac{18442465574893}{5715360000} \\
		\frac{852816554717}{11430720000} & \frac{13030989991}{2286144000} \\
	\end{array}
	\right),
\end{eqnarray*}
\begin{eqnarray*}
	\textbf{B}^{u}(8)&=&\left(
	\begin{array}{ccccc}
		0 & -\frac{872897054659901}{27461161728} & -\frac{18588146435171828297}{70025962406400} & -\frac{413569750046670214409}{350129812032000} & -\frac{14787600350917688774251}{4376622650400000} \\
		0 & -\frac{950052494454097}{11442150720} & -\frac{13244337545102475247}{8581613040000} & -\frac{264367876414238436491}{21454032600000} & -\frac{301580561206259400151}{6129723600000} \\
		0 & -\frac{16238352579201839}{68652904320} & -\frac{2373252477645589592807}{625231807200000} & -\frac{96025127809245373230077}{3501298120320000} & -\frac{278085466866232929313049}{2693306246400000} \\
		0 & -\frac{13143770575741441}{68652904320} & -\frac{995937783105265674239}{257448391200000} & -\frac{2350421651369022799319}{73556683200000} & -\frac{5331785324269380365833}{41191742592000} \\
		0 & -\frac{9444733487400923}{68652904320} & -\frac{558926136337582161637}{257448391200000} & -\frac{1589239543081246708199}{102979356480000} & -\frac{59148943373367455404739}{1029793564800000} \\
		0 & -\frac{507339953516177}{13730580864} & -\frac{136657625883731073467}{257448391200000} & -\frac{1996276255302159037997}{514896782400000} & -\frac{1416193691744932910423}{93617596800000} \\
		0 & -\frac{11678789233897}{980755776} & -\frac{45265835304985762657}{437662265040000} & -\frac{1666765675235287627447}{3501298120320000} & -\frac{48614865808330901615719}{35012981203200000} \\
		0 & -\frac{53092182089}{39007332} & -\frac{21947335344624203}{2925549900000} & -\frac{454524239184255743}{23404399200000} & -\frac{338045490925866803}{9361759680000} \\
		0 & -\frac{1666893622223}{4992938496} & -\frac{34769068111387841513}{17506490601600000} & -\frac{16813066682918948423}{3501298120320000} & -\frac{1295331946972830623}{218831132520000} \\
	\end{array}
	\ldots\ldots\right.\\
	&&\left.\ldots\ldots
	\begin{array}{cccc}
		-\frac{26671980610761545640463}{4376622650400000} & -\frac{10410924449492715109187}{1591499145600000} & -\frac{33111769841940340290731}{8753245300800000} & -\frac{15828728653296464701673}{17506490601600000} \\
		-\frac{456246190744402203133}{4290806520000} & -\frac{217327134904346719957}{1716322608000} & -\frac{1677198534587955706049}{21454032600000} & -\frac{838693437524617003943}{42908065200000} \\
		-\frac{47035289606202568606931}{218831132520000} & -\frac{1745666266438961774155609}{7002596240640000} & -\frac{105629267865011237851279}{700259624064000} & -\frac{129929631322226023742719}{3501298120320000} \\
		-\frac{1813232483719265888827}{6436209780000} & -\frac{346809455737849275209449}{1029793564800000} & -\frac{107297330482079718457187}{514896782400000} & -\frac{5372707806003957287993}{102979356480000} \\
		-\frac{6109681701556168557649}{51489678240000} & -\frac{2877194638779880164413}{21016195200000} & -\frac{8498901170351000106563}{102979356480000} & -\frac{5212373807745931214239}{257448391200000} \\
		-\frac{4190072201435437478723}{128724195600000} & -\frac{39941538353675711318161}{1029793564800000} & -\frac{2470393487607124610383}{102979356480000} & -\frac{281309819466721802633}{46808798400000} \\
		-\frac{22145687272058418767953}{8753245300800000} & -\frac{2736808376419680966983}{1000370891520000} & -\frac{27906138187108288734971}{17506490601600000} & -\frac{3363785027662396568069}{8753245300800000} \\
		-\frac{1282549052137542647}{23404399200000} & -\frac{682401984740784199}{11702199600000} & -\frac{410522570657836837}{11702199600000} & -\frac{58422103525040867}{6686971200000} \\
		-\frac{4814954013850918313}{1250463614400000} & -\frac{43424519219943077927}{35012981203200000} & -\frac{2889033335083336057}{17506490601600000} & -\frac{200810348470291303}{35012981203200000} \\
	\end{array}
	\right)
\end{eqnarray*}
, and the coefficient of lower boundary is $\beta_{d}^{(k)}(\alpha,n)=\sum_{i=1}^{k+1}\sum_{j=1}^{k+1}\alpha^{i-1} B^{d}_{i,j}(k)n^{j-1}$ with the coefficient matrix $B^{d}_{i,j}(k)$ are
\begin{eqnarray*}
	\textbf{B}^{d}(1)&=&\left(
	\begin{array}{cc}
		0 & 4 \\
		0 & 2 \\
	\end{array}
	\right)\hspace{10px}
	\textbf{B}^{d}(2)=\left(
	\begin{array}{ccc}
		-2 & 4 & 6 \\
		0 & 16 & 16 \\
		-1 & -2 & -1 \\
	\end{array}
	\right),\hspace{10px}
	\textbf{B}^{d}(3)=\left(
	\begin{array}{cccc}
		0 & 28 & 72 & 44 \\
		0 & 74 & 216 & 142 \\
		0 & 26 & 72 & 46 \\
		0 & 1 & 0 & -1 \\
	\end{array}
	\right),\\
	\textbf{B}^{d}(4)&=&\left(
	\begin{array}{ccccc}
		-\frac{43}{30} & \frac{97}{10} & \frac{7969}{60} & \frac{2983}{10} & \frac{10597}{60} \\
		0 & 308 & 1814 & 3020 & 1514 \\
		-\frac{14}{5} & \frac{558}{5} & \frac{11603}{15} & \frac{7042}{5} & \frac{11239}{15} \\
		0 & 28 & \frac{518}{3} & 292 & \frac{442}{3} \\
		-\frac{1}{60} & -\frac{41}{20} & -\frac{977}{120} & -\frac{159}{20} & -\frac{221}{120} \\
	\end{array}
	\right),
	\textbf{B}^{d}(5)=\left(
	\begin{array}{cccccc}
		0 & \frac{45938}{225} & \frac{136048}{75} & \frac{2303461}{450} & \frac{420152}{75} & \frac{941863}{450} \\
		0 & \frac{531083}{450} & \frac{1125787}{100} & \frac{59852027}{1800} & \frac{3846213}{100} & \frac{27519641}{1800} \\
		0 & \frac{33296}{25} & \frac{2777957}{225} & \frac{16131283}{450} & \frac{9199243}{225} & \frac{7223789}{450} \\
		0 & \frac{35314}{75} & \frac{972896}{225} & \frac{2810887}{225} & \frac{3189904}{225} & \frac{1245971}{225} \\
		0 & \frac{13361}{450} & \frac{29329}{100} & \frac{541003}{600} & \frac{314413}{300} & \frac{737947}{1800} \\
		0 & -\frac{3769}{900} & -\frac{19769}{1800} & -\frac{4087}{1200} & \frac{569}{1800} & -\frac{11063}{3600} \\
	\end{array}
	\right),\\
	\textbf{B}^{d}(6)&=&\left(
	\begin{array}{ccccccc}
		-\frac{3161}{1500} & \frac{26927833}{108000} & \frac{4598004583}{1512000} & \frac{5795932571}{378000} & \frac{1855393303}{56000} & \frac{24357112027}{756000} & \frac{4396126631}{378000} \\
		0 & \frac{6898699}{1575} & \frac{968634697}{15750} & \frac{8855842207}{31500} & \frac{105342988399}{189000} & \frac{15755047813}{31500} & \frac{31526579237}{189000} \\
		-\frac{3703}{300} & \frac{403751777}{108000} & \frac{37990647151}{648000} & \frac{30799301317}{108000} & \frac{47795705483}{81000} & \frac{3286459717}{6000} & \frac{24361535093}{129600} \\
		0 & \frac{297424051}{75600} & \frac{40853120603}{756000} & \frac{3436866121}{14000} & \frac{11464395464}{23625} & \frac{82144827239}{189000} & \frac{36476848183}{252000} \\
		-\frac{3179}{1500} & \frac{305626919}{756000} & \frac{1107731089}{181440} & \frac{4560915011}{151200} & \frac{10265293207}{162000} & \frac{7536347671}{126000} & \frac{3794314171}{181440} \\
		0 & \frac{46649}{1575} & \frac{17643341}{47250} & \frac{29754191}{15750} & \frac{364325537}{94500} & \frac{18243773}{5250} & \frac{12233309}{10500} \\
		\frac{3}{500} & \frac{5949379}{504000} & \frac{484129979}{9072000} & \frac{2493949}{31500} & \frac{161105327}{9072000} & -\frac{10236563}{504000} & -\frac{2100869}{4536000} \\
	\end{array}
	\right),
\end{eqnarray*}
\begin{eqnarray*}
	\textbf{B}^{d}(7)&=&\left(
	\begin{array}{cccccc}
		0 & \frac{397333711}{236250} & \frac{912481651201}{31752000} & \frac{173427383585083}{952560000} & \frac{799148282837}{1512000} & \frac{183712783847737}{238140000} \\
		0 & \frac{15055932377}{945000} & \frac{19265448332759}{63504000} & \frac{530128616599171}{272160000} & \frac{202315617785353}{35280000} & \frac{8115116669833181}{952560000} \\
		0 & \frac{537797233}{16875} & \frac{57011835834479}{95256000} & \frac{10827125498380181}{2857680000} & \frac{5250459223599319}{476280000} & \frac{11578483272574709}{714420000} \\
		0 & \frac{292847801}{11250} & \frac{93962755567787}{190512000} & \frac{17995590164805689}{5715360000} & \frac{8785533736833223}{952560000} & \frac{9743281244777473}{714420000} \\
		0 & \frac{4326178303}{472500} & \frac{8108389258903}{47628000} & \frac{1547327421338977}{1428840000} & \frac{754417201616777}{238140000} & \frac{3344755232825411}{714420000} \\
		0 & \frac{8685673}{7500} & \frac{4079789421839}{190512000} & \frac{793436945079797}{5715360000} & \frac{55667390342629}{136080000} & \frac{107916030493711}{178605000} \\
		0 & \frac{27559079}{945000} & \frac{11827172977}{27216000} & \frac{3547898619617}{1143072000} & \frac{9219646684781}{952560000} & \frac{8647995409723}{571536000} \\
		0 & -\frac{279383}{70000} & -\frac{2224959083}{54432000} & -\frac{1621977063431}{11430720000} & -\frac{421769056267}{1905120000} & -\frac{434184850123}{5715360000} \\
	\end{array}
	\ldots\ldots\right.\\
	&&\left.\ldots\ldots
	\begin{array}{cc}
		\frac{8711068988611}{15876000} & \frac{20517348363031}{136080000} \\
		\frac{493432968607007}{79380000} & \frac{3373911641427409}{1905120000} \\
		\frac{398276643093449}{34020000} & \frac{9436217113175831}{2857680000} \\
		\frac{4718411159783921}{476280000} & \frac{3212480565675379}{1143072000} \\
		\frac{28904032500311}{8505000} & \frac{1375712097341929}{1428840000} \\
		\frac{207224725526201}{476280000} & \frac{99107423249021}{816480000} \\
		\frac{338682310657}{29767500} & \frac{18442465574893}{5715360000} \\
		\frac{16530496043}{476280000} & -\frac{13030989991}{2286144000} \\
	\end{array}
	\right),\\
	\textbf{B}^{d}(8)&=&\left(
	\begin{array}{ccccc}
		-\frac{315276383}{64260000} & \frac{2788741300997412661}{972582811200000} & \frac{1367043564877117357}{25261891200000} & \frac{83668627281908448049}{182359277100000} & \frac{33700440149666544069539}{17506490601600000} \\
		0 & \frac{43765104774011}{764032500} & \frac{280242187907003197}{198648450000} & \frac{5389618946891445769}{446959012500} & \frac{1050536592960008240431}{21454032600000} \\
		-\frac{3933833861}{80325000} & \frac{172012315478400804109}{1945165622400000} & \frac{817038838348186882951}{353666476800000} & \frac{59977587466719441052789}{2917748433600000} & \frac{1505686264504828318680281}{17506490601600000} \\
		0 & \frac{44419931204714423}{293388480000} & \frac{47827258839539894509}{12713500800000} & \frac{395154853935328437281}{12259447200000} & \frac{13493281265625073581797}{102979356480000} \\
		-\frac{9702061}{315000} & \frac{206234437563110213}{4400827200000} & \frac{4033391679143310347}{3269185920000} & \frac{473921127831031617683}{42908065200000} & \frac{2985505579185826448329}{64362097800000} \\
		0 & \frac{2368934530074779}{125737920000} & \frac{7290035098296442379}{16345929600000} & \frac{323411483779967104817}{85816130400000} & \frac{7838859586585197529459}{514896782400000} \\
		-\frac{16082767}{11475000} & \frac{592152467309470673}{389033124480000} & \frac{22011497665815279697}{778066248960000} & \frac{12481333639755250253}{56110546800000} & \frac{7738913158846465383521}{8753245300800000} \\
		0 & \frac{2272417081}{69457500} & \frac{58682998912897}{108353700000} & \frac{36890627729928373}{7801466400000} & \frac{192505370574765427}{9361759680000} \\
		\frac{104597}{91800000} & -\frac{841786729152769}{77806624896000} & -\frac{61448270893941083}{555761606400000} & -\frac{102103272786152767}{364718554200000} & -\frac{729654237448525603}{7002596240640000} \\
	\end{array}
	\ldots\ldots\right.\\
	&&\left.\ldots\ldots
	\begin{array}{cccc}
		\frac{6345630203881728810623}{1458874216800000} & \frac{7243061034581239521359}{1346653123200000} & \frac{1438244989106929453141}{416821204800000} & \frac{15828728653296464701673}{17506490601600000} \\
		\frac{152036745896413284299}{1430268840000} & \frac{494165012642414929513}{3900733200000} & \frac{559191738503504103241}{7151344200000} & \frac{838693437524617003943}{42908065200000} \\
		\frac{2231401312787071933213}{11670993734400} & \frac{1627676869932832454158343}{7002596240640000} & \frac{15493657916750060626829}{106099943040000} & \frac{129929631322226023742719}{3501298120320000} \\
		\frac{488280046911468358531}{1716322608000} & \frac{349005014669841257446871}{1029793564800000} & \frac{5124332464670408241073}{24518894400000} & \frac{5372707806003957287993}{102979356480000} \\
		\frac{8875339744294997232979}{85816130400000} & \frac{3709866338376811902617}{29422673280000} & \frac{13634491690726632965003}{171632260800000} & \frac{5212373807745931214239}{257448391200000} \\
		\frac{706660785334913765471}{21454032600000} & \frac{3663713001152510819389}{93617596800000} & \frac{4134432225011965193263}{171632260800000} & \frac{281309819466721802633}{46808798400000} \\
		\frac{511947365891879033131}{265249857600000} & \frac{11692325950480172454077}{5001854457600000} & \frac{8638140751830018784711}{5835496867200000} & \frac{3363785027662396568069}{8753245300800000} \\
		\frac{371686362676910173}{7801466400000} & \frac{671427062862928823}{11702199600000} & \frac{135795626230911767}{3900733200000} & \frac{58422103525040867}{6686971200000} \\
		\frac{1302865197722412781}{2917748433600000} & \frac{8600742285944529613}{35012981203200000} & -\frac{19864685154306389}{166728481920000} & \frac{200810348470291303}{35012981203200000} \\
	\end{array}
	\right).
\end{eqnarray*}

\end{widetext}

\clearpage
\newpage

\clearpage
\newpage
\begin{figure}[t]
	\centering
	\includegraphics[width=0.5\linewidth ]{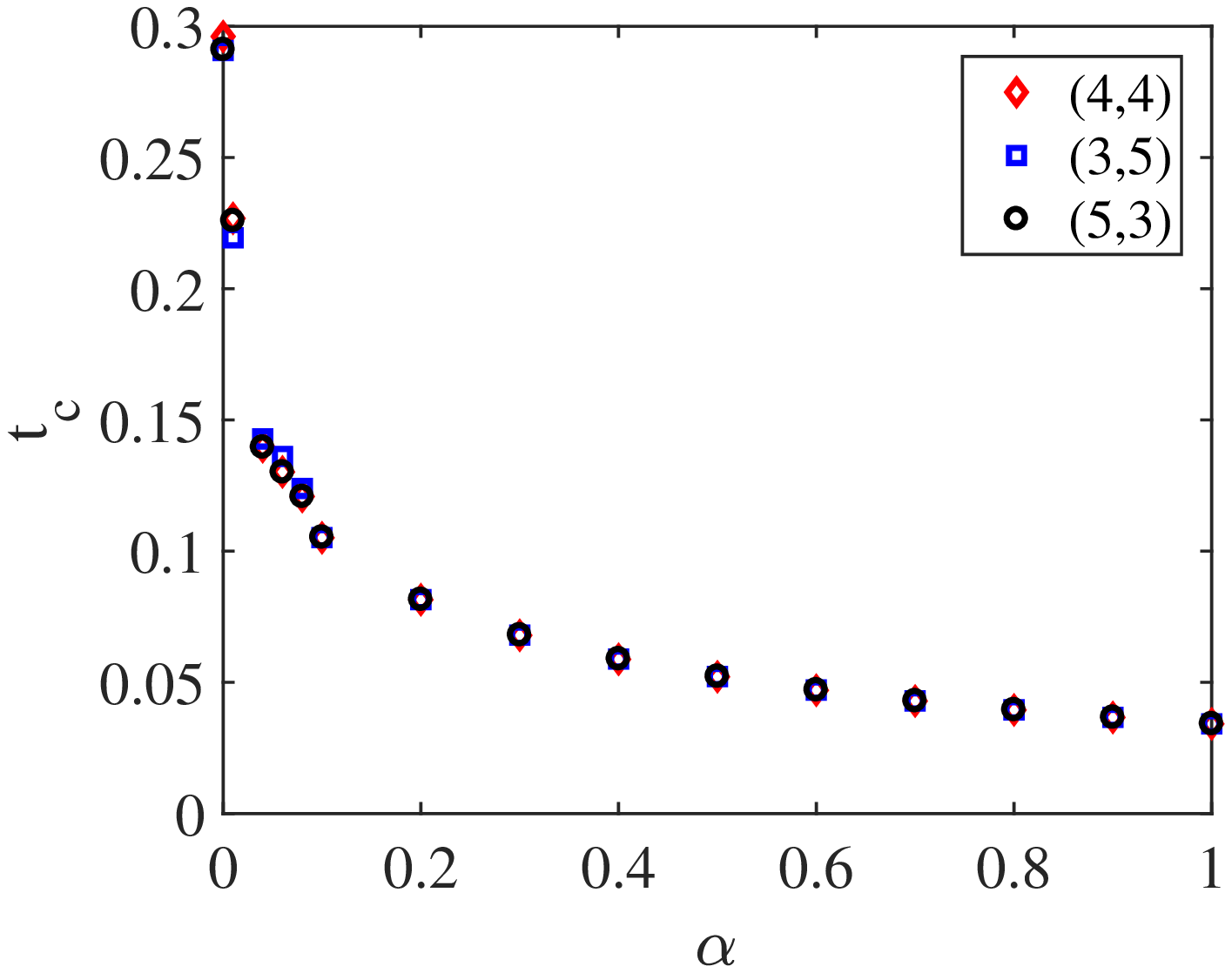}\hfill
	\includegraphics[width=0.5\linewidth ]{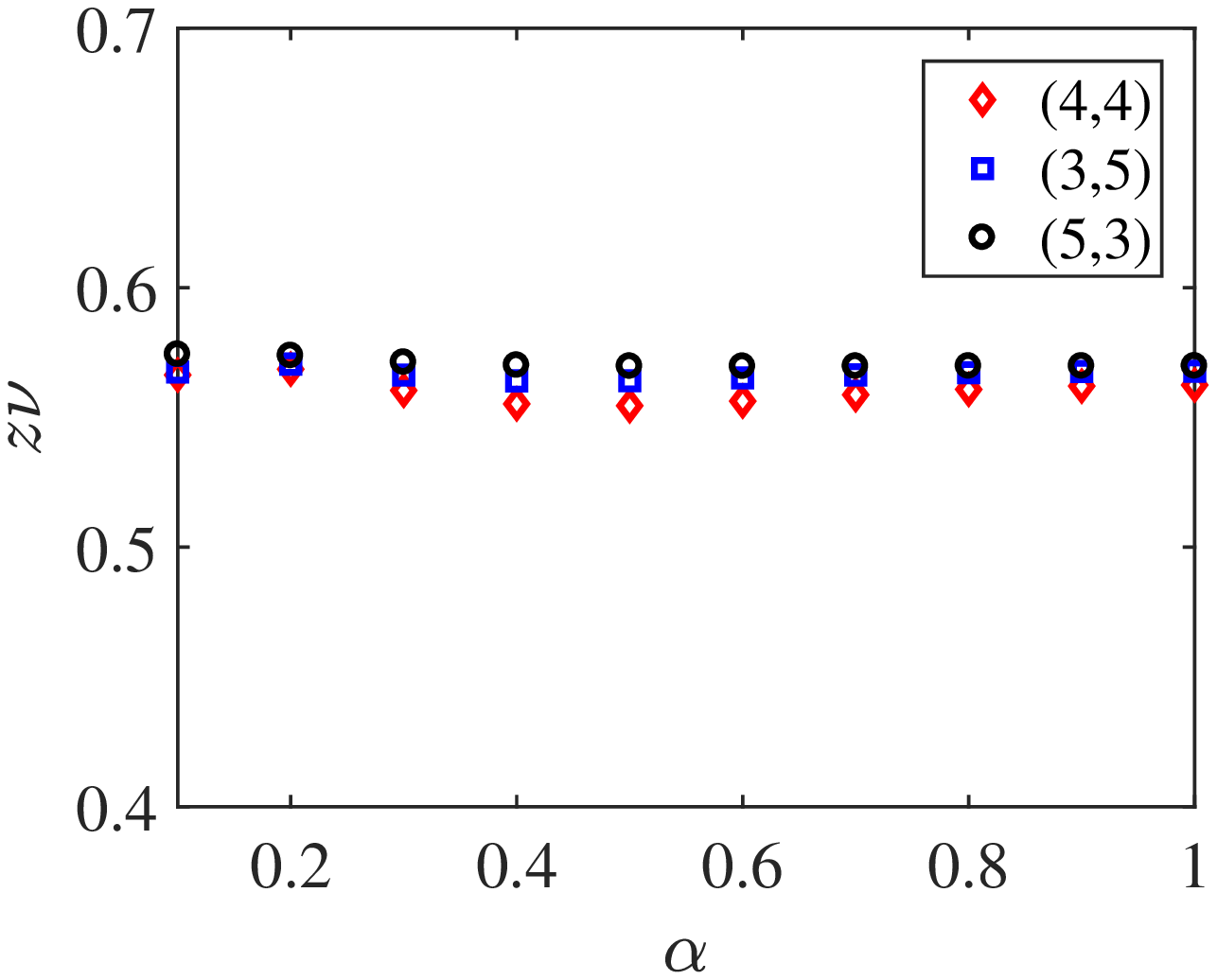}\hfill
	\caption{The critical hoppings and exponents for different P\'{a}de degrees at same commensurate filling.}\label{s1}
\end{figure}

\section{Sensitivity of Pade resummation}\label{AA2}
A key step for us to get the critical hopping amplitude and the corresponding critical exponent $z\nu$ from our eighth order strong coupling expansion series is using P\'{a}de re-summation method as following:
\begin{equation}
f(t)=\frac{\sum_{n=0}^{n_{max}}a_nt^n}{1+\sum_{m=0}^{m_{max}}b_mt^m}.\label{pade2}
\end{equation} 
The coefficients $a_n$ and $b_m$ could be calculated by fitting with the restriction $n_{max}+m_{max}=8$. The most nature way to choose the parameters $(n_{max},m_{max})$ is $(4,4)$. In order to check the effect of P\'{a}de degrees chosen, we compare $(4,4)$ results with the case of $(5,3)$ and $(3,5)$. As shown in Fig.\ref{s1}, the small discrepancy of both critical hopping amplitude and critical exponents demonstrates P\'{a}de degrees has less influence.

\end{appendix}

\end{document}